\documentclass[11pt]{article}

\usepackage[margin=1in]{geometry}
\usepackage{graphicx,amsmath,amsthm,amssymb,mathtools,hyperref,enumerate,verbatim,soul,tikz,cite}

\newtheorem{theorem}{Theorem}

\newtheorem{lemma}[theorem]{Lemma}

\newtheorem{definition}[theorem]{Definition}

\newtheorem{question}[theorem]{Question}

\numberwithin{theorem}{section}
\numberwithin{equation}{section}

\title{Topological Big Bangs: \\ Reflection, Itty-Bitty Blenders, and Eternal Trumpets}

\author{Hubert Bray\footnote{Department of Mathematics, Duke University, Durham, NC 27708, USA. bray@math.duke.edu.} \and James Wheeler\footnote{Department of Mathematics, University of Michigan, Ann Arbor, MI 48109, USA. jcwheel@umich.edu.} \footnote{Corresponding author.}}

\date{}

\begin{document}

\maketitle

\begin{abstract} 
We discuss and formalize topological means by which the initial singularity might be mollified, at the level of the spacetime manifold's structure, in classical cosmological models of a homogeneous expanding universe. 
One construction, dubbed a ``reflective" {\it topological big bang}, generalizes Schr\"odinger's elliptic de Sitter space and is built to be compatible with the standard Friedmann-Lema\^itre-Robertson-Walker (FLRW) picture of the large-scale universe, only minimally modifying it via some nontrivial topology at an earliest ``moment" in the universe's history. 
We establish a mathematical characterization of the admissible topological structures of reflective topological big bangs, and we discuss implications for a standard concern in cosmology, the horizon problem. 
We present a nonreflective example that we've christened the Itty-Bitty Blender spacetime: this spacetime and its universal cover, the Eternal Trumpet spacetime, exhibit interesting potential structures of spacetimes avoiding the Hawking and Penrose singularity theorems. While these toy models provide a proof-of-concept picture, several questions remain regarding the capacity to realize these structures under physical energy conditions.
\end{abstract} 

\section{Introduction}

While the essential features of the big bang cosmological model, as realized in general relativity through the homogeoneous and isotropic Friedmann-Lema\^itre-Robertson-Walker (FLRW) spacetimes and perturbations thereof, have been remarkably successful in explaining a wide array of observable phenomena, physicists and cosmologists are broadly interested in avenues by which the initial singularity might be eliminated. 
Whatever the final theory of cosmology should be, physicists generally anticipate that the singularity will give way to something calculable. 
As quantum effects should increasingly dominate as one traces the universe's expansion backwards, it is widely held that the singularity's full resolution will be inextricably tied to a theory of quantum gravity. 
It may end up that the nature of this theory will render the classical picture of spacetime moot in this regime, but it is also possible that a manifold structure of spacetime continues to be meaningful, and one might then ask what kind of structure the manifold could have. 

A variety of possibilities for the nature of the early universe have been considered, with the most prominent propositions involving either inflation (very much the standard paradigm)-- early references include \cite{guth1981inflationary,linde1982new,albrecht1982cosmology,goncharov1987global}; see \cite{achucarro2022inflation} for a recent review of the theoretical and observational status of inflation-- or (perhaps cyclic) ``bouncing" cosmologies-- reviews include \cite{battefeld2015critical,brandenberger2017bouncing,ijjas2018bouncing}.
See also \cite{misner1969mixmaster,starobinsky1980new, vilenkin1983birth}. 
In this work we present, at the classical level, a few plausible structures of the spacetime manifold naturally compatible with FLRW metrics that morally amount to replacing the singularity with some nontrivial topology appended early in the universe's evolution. 
Although they could in principle, these need not be thought of as supplanting inflation. 
Indeed, as typical inflationary models are thought to be past-incomplete \cite{borde1996singularities,borde2003inflationary,guth2007eternal}, they leave open the nature of the ``initial" structure of the universe. Even though the various models we present below are constructed to connect directly to FLRW geometries, then, we view them as propositions for plausible topological structures of the early universe that are independent of the existence of an inflationary era-- such an era may or may not occur ``after" these structures. That is, these are essentially variations on the FLRW {\it framework} within which one might try to implement reasonable physics, rather than specific instantiations of reasonable physics. Of course, if these frameworks are to convincingly operate as directly suggested by the given toy models, without a mechanism such as inflation, they must afford implementations of reasonable physics reproducing the various successes of inflation, such as resolving the horizon and flatness problems and generating a nearly scale-invariant power spectrum of primordial curvature perturbations. We do not claim to supply such implementations in this proof-of-concept work, and hence we also cannot currently speculate on unique observable signatures.  We do, however, consider interplay of each model with the horizon problem, which is somewhat uniquely amenable to discussion at this level of breadth.

The remainder of this work is organized as follows. In Section \ref{sec:reflective}, we introduce, define, and characterize ``reflective" topological big bangs: We classify the possible orientable manifold structures compatible with them in Section \ref{sec:classify}, and we discuss their relation to the horizon problem as a means of illustrating their geometry in Section \ref{sec:horizon}. These examples are not time-orientable, but this pathology is, in a sense, entirely localized to the ``beginning" of the universe.
In Section \ref{subsec:ittybitty}, we discuss an example, dubbed the Itty-Bitty Blender spacetime, of what one might call a topological big bang that is not reflective. 
This example is time-orientable, but it admits closed timelike curves also localized at the ``beginning" of the universe. 
In Section \ref{subsec:trumpet}, however, we discuss the Eternal Trumpet spacetime, the blender's topologically trivial universal cover, which is globally hyperbolic. 
We summarize and conclude in Section \ref{sec:conclusion}.

Each spacetime explicitly written down in this work modeling a topological big bang violates standard energy conditions near the new structures appended to an FLRW spacetime. One may take the perspective that this is not particularly concerning since the actual matter in these regimes may well require a quantum mechanical description (perhaps depending on how early one appends the structures in question), rendering the pointwise imposition of energy conditions questionable. We consider this the best resolution of this concern for reflective topological big bangs. Precisely what makes the blender and trumpet examples interesting, however, is that their essential features do not {\it necessitate} the violation of physical energy conditions according to known theorems. Much of the discussion of Section \ref{subsec:trumpet} is devoted to these examples' interplay with Hawking and Penrose's celebrated singularity theorems \cite{penrosesingularity,hawking1966occurrence,hawking1966occurrence2,hawking1967occurrence,hawking1970singularities} in cosmology, leading to several interesting open questions as to whether their structures can be replicated via more physically compelling metrics.

\section{Reflective Topological Big Bangs}
\label{sec:reflective}

Though we are generally interested in orientable spacetimes, we begin with a nonorientable (in $n=4$) example because it admits a simple and illustrative picture, and because it arises from a common toy model of cosmology: de Sitter space.
We review the well-known construction of ``elliptic" de Sitter space, originally considered by Schr\"odinger in a 1956 monograph \cite{schrodinger2011expanding}; more recently, it has been studied as a potentially natural backdrop for quantum field theory \cite{sanchez1987quantum,parikh2003elliptic,halpern2015holography}, and as a means of avoiding past-incompleteness theorems relevant to inflation \cite{aguirre2003inflation}.

\begin{figure}[t]
\centering
\includegraphics[width=\textwidth]{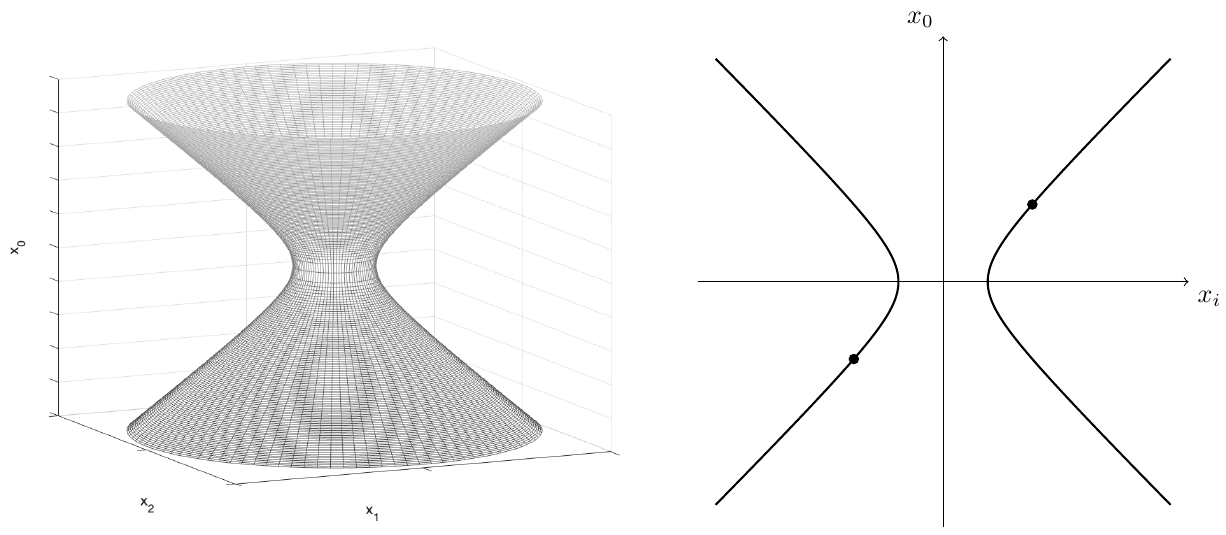}
\caption{\small De Sitter space as a submanifold of Minkowski space for $n = 2$. On the right, a pair of points to be identified is shown in a vertical cross-section.}
\label{fig:hyperboloid}
\end{figure}

We recall that a standard definition of $n$-dimensional de Sitter space identifies it as a submanifold embedded in $\mathbb R^{1,n}$, $(n+1)$-dimensional Minkowski space. We endow $\mathbb R^{1,n}$ with coordinates $x_0,x_1,\dots,x_n$ with respect to which the standard metric is
\begin{equation} \label{eqn:minkowski}
ds^2 = -dx_0^2 + \sum_{i = 1}^n dx_i^2.
\end{equation}
Given $\alpha \in \mathbb R$, a de Sitter space with constant sectional curvature $-\alpha^{-2}$ is realized as the locus of
$$
-x_0^2 + \sum_{i=1}^n x_i^2 = \alpha^2,
$$
a hyperboloid (see Figure \ref{fig:hyperboloid}), with a Lorentzian metric induced by (\ref{eqn:minkowski}) and topology $\mathbb R \times S^{n-1}$. This hyperboloid can be foliated a number of ways to yield different-appearing FLRW metrics with Euclidean, hyperbolic, or spherical spatial slices. We draw on the most straightforward picture given the Minkowski embedding, wherein the spatial slices of constant time are level sets of $x_0$, i.e.\@ $(n-1)$-spheres of radius $\alpha^2 + x_0^2$. In particular, defining $t$ by $x_0 = \alpha \sinh(t/\alpha)$, the hyperboloid's induced metric takes the form
$$
ds^2 = - dt^2 + \alpha^2 \cosh^2(t/\alpha) d\Omega^2_{n-1},
$$
where $d\Omega^2_{n-1}$ is the standard metric on the unit $(n-1)$-sphere.

\begin{figure}[t]
\centering
\includegraphics[width=0.5\textwidth]{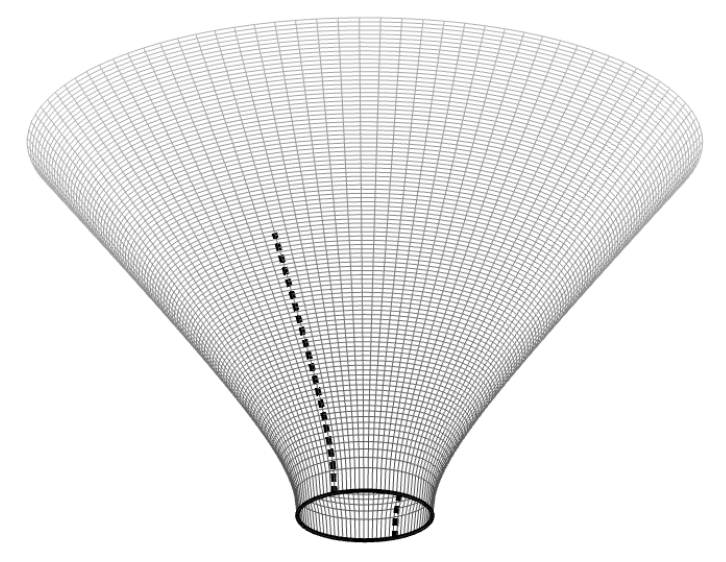}
\caption{\small The quotient of de Sitter space obtained via antipodal identification. The dotted line is an example of a continuous curve that passes through the reflection surface, emerging on the opposite side.}
\label{fig:quotient}
\end{figure}

Under this foliation, the universe apparently ``bounces" from contraction to expansion at $t = 0$, offering a schematic illustration of one possible mollification of the singularity. 
To obtain a qualitatively different mollification, one may modify de Sitter space by quotienting by antipodal identification, where antipodes $p \mapsto -p$ are taken in the ambient Minkowski space: this is elliptic de Sitter space, which we now denote by $M$. The contraction prior to the bounce and expansion thereafter are now identified, so that the hyperboloid is effectively reduced to a funnel shape, illustrated in Figure \ref{fig:quotient}. The ``mouth" of the funnel $M_0$ (the quotiented image of $\{0\} \times S^{n-1}$), acting as a sort of reflection surface, is now an $\mathbb{RP}^{n-1}$ instead of an $\mathbb S^{n-1}$, and the metric remains perfectly smooth across it since the antipodal identification was isometric. Due to this identification, a continuous curve which passes through the mouth apparently emerges on the opposite end of the spatial universe, as these opposite ends are smoothly joined along $M_0$. 

As de Sitter space provides a bare-bones picture of a bouncing cosmology, elliptic de Sitter space provides a bare-bones picture of a nonsingular expanding universe which extends only finitely to the past, in what we dub a ``reflective" topological big bang, defined formally in Definition \ref{def:topbigbang} below. This spacetime, like all {\it reflective} topological big bangs, has the pathology that it is not time-orientable: a curve passing through $M_0$ and returning to its starting position reverses time orientation. This pathology, however, is in a sense entirely localized at the reflection surface $M_0$, thought of as extremely close to the standard cosmology's would-be singularity. $M \backslash M_0$ remains time-orientable, and indeed globally hyperbolic: one can still consider that $(\epsilon, \infty) \times S^{n-1} \subset M$ is the future Cauchy development of the spatial slice $\Sigma_\epsilon := \{\epsilon\} \times S^{n-1}$, so an initial value problem initialized in the classical regime remains meaningful. Schematically, then, we've replaced the complete pathology of an initial singularity with a lesser pathology in initial causality, which may (or may not) be compatible with an eventual picture of quantum gravity.

While elliptic de Sitter space is the best image to keep in mind, it has the perhaps less palatable pathology in $n=4$ (and indeed any even dimension) that it is nonorientable.
In particular, this spacetime cannot admit continuous spinor fields \cite{geroch1968spinor}, seemingly a most basic requirement if general relativity is to ultimately connect to quantum field theory. 
Although working reflective topological big bangs will always be closely associated with a symmetric bouncing cosmology (note that, due to the issue of orientability, the reverse is not true) obtainable by passing to a covering space, these provide qualitatively different pictures of the universe's structure, so it is interesting to consider them separately and enumerate the possible ways of constructing a reflective topological big bang.

\subsection{Classifying Reflective Topological Big Bangs}
\label{sec:classify}

The defining characteristic distinguishing a reflective topological big bang model from a standard FLRW topology $M \cong \mathbb R_+ \times \Sigma$ (for some $(n-1)$-manifold $\Sigma$) is the reflection surface, an ``earliest" spatial slice that is {\it one-sided} as an embedded submanifold. Away from this surface, our spacetime should be identical to an FLRW spacetime. The following definition provides this picture, at the topological level, in the most mathematically streamlined fashion:

\begin{definition} \label{def:topbigbang}
    A smooth $n$-manifold $M$ is said to model a \ul{reflective topological big bang} if it is diffeomorphic to $(\mathbb R \times \Sigma)/\mathbb Z_2$ for some connected $(n-1)$-manifold $\Sigma$, where the $\mathbb Z_2$ action is defined by $(t,p) \mapsto (-t, \sigma(p))$ for each $(t,p) \in \mathbb R \times \Sigma$, where $\sigma : \Sigma \to \Sigma$ is a smooth free involution. \\ In such a model, the \underline{reflection surface} is $M_0 := (\{0\} \times \Sigma)/\mathbb Z_2 \cong \Sigma / \sigma$.
\end{definition}

 It is straightforward to deduce that $M_0$ must be one-sided:

\begin{lemma}
Let $M \cong (\mathbb R \times \Sigma)/\mathbb Z_2$ model a reflective topological big bang. The reflection surface $M_0 \subset M$ is one-sided, and $\Sigma$ is a double cover of $M_0$.
\end{lemma}

\proof That $\Sigma$ is a double cover of $M_0$ is immediate from Definition \ref{def:topbigbang}, so we establish that $M_0$ is one-sided. Any tubular neighborhood of $M_0$ in $M$ lifts to a neighborhood of $\{0\} \times \Sigma$ in $\mathbb R \times \Sigma$, and the tube lemma indicates that this contains an open set of the form $(-\epsilon,\epsilon) \times \Sigma$. Projecting back to $M$, observe that $((-\epsilon,\epsilon) \times \Sigma)/\mathbb Z_2 \subset M$ remains connected upon removing $M_0$. If the normal bundle $N M_0$ were trivial, however, no neighborhood of the zero section could have this property, so $M_0$ must be one-sided.
\\ \qed

It is well-known that if a hypersurface is embedded in an orientable manifold, being one-sided is equivalent to being nonorientabile. As every nonorientable manifold admits a unique oriented double cover (the orientation double cover; see Theorem 15.42 of \cite{lee2003smooth}), this leads to our essential result, formalized below in Theorem \ref{thm:classification}, that the nonorientable reflection surface $M_0$ precisely and explicitly characterizes the topological structure of any orientable $M$ modeling a reflective topological big bang.

\begin{theorem} \label{thm:classification}
Given any smooth, nonorientable, and connected $(n-1)$-manifold $\widetilde \Sigma$, there exists a unique orientable $n$-manifold $M$ modeling a reflective topological big bang into which $\widetilde \Sigma$ embeds as the reflection surface $M_0$, constructible explicitly as $M = \bigwedge^{n-1} T^* \widetilde \Sigma$, the line bundle of top forms over $\widetilde \Sigma$. The spatial slice surface $\Sigma$ in this model is the orientation double cover of $\widetilde \Sigma$.
\end{theorem}

\proof Take $M = \bigwedge^{n-1} T^* \widetilde \Sigma$, and note that $\widetilde \Sigma$ naturally embeds into $M$ as the zero section $M_0$. Endowing $\widetilde \Sigma$ with some smooth Riemannian metric $\tilde g$, define $t : M \to \mathbb R$ by $t(\omega_p) = \|\omega_p\|_{\tilde g}$ for each top form $\omega_p \in M$ based at $p \in \widetilde \Sigma$. We observe that $M_+ := t^{-1}(\mathbb R_+)$ is an open submanifold. As each $t_0 > 0$ is a regular value of $t$, the level sets $M_{t_0} := t^{-1}(t_0)$ foliating $M_+$ are mutually diffeomorphic embedded submanifolds: denoting the hypersurface of unit top forms by $\Sigma := M_1$, $M_+$ is apparently diffeomorphic to $\mathbb R_+ \times \Sigma$. 

Identifying each $\omega_p \in M$ with an orientation of $T_p \widetilde \Sigma$ in the usual way, it is straightforward to see that $\Sigma = M_1$ is diffeomorphic to the orientation double cover of $\widetilde \Sigma$ (with the covering map being given by restricting the bundle projection $\pi: M \to \widetilde \Sigma$ to $\Sigma$). In particular, $\Sigma$ is connected and orientable, as is $\mathbb R \times \Sigma$, and the smooth free involution $\sigma : \Sigma \to \Sigma$ given by $\sigma(\omega_p) = - \omega_p$ has the property that the induced $\mathbb Z_2$ action on $\mathbb R \times \Sigma$ (as in Definition \ref{def:topbigbang}) is orientation-preserving, so that $(\mathbb R \times \Sigma)/ \mathbb Z_2$ is orientable. Finally, that the $\mathbb R \times \Sigma \to M$ map 
$(s,\omega_p) \mapsto s \omega_p$ is a local diffeomorphism\footnote{Given $\omega_p \in \Sigma$, this is the identity map in appropriate coordinates on $\mathbb R \times U$ and $\pi^{-1}(\widetilde U) \subset M$, with $U \subset \Sigma$ a covering sheet containing $\omega_p$ of a sufficiently small $M$-trivializing coordinate neighborhood $\widetilde U \subset \widetilde \Sigma$ of $p$.}, respects the $\mathbb Z_2$ action, and becomes a bijection upon quotienting establishes that $M \cong (\mathbb R \times \Sigma)/ \mathbb Z_2$. Under this diffeomorphism, the zero section $M_0 \cong \widetilde \Sigma$ is identified with the reflection surface $(\{0\} \times \Sigma)/ \mathbb Z_2$.

Now suppose $\widehat M \cong (\mathbb R \times \widehat \Sigma)/ \mathbb Z_2 $ is another orientable $n$-manifold modeling a reflective topological big bang with reflection surface $\widetilde \Sigma \cong (\{0\} \times \widehat \Sigma)/ \mathbb Z_2 \cong \widehat \Sigma/\widehat \sigma$. As $\widehat \Sigma$ embeds into $\widehat M$ as a two-sided hypersurface (e.g.\@ $(\{\pm 1\} \times \widehat \Sigma)/\mathbb Z_2$) and is a double cover of $\widetilde \Sigma$, $\widehat \Sigma$ must be diffeomorphic to the orientation double cover of $\widetilde \Sigma$, so $\widehat \Sigma \cong \Sigma$ in a manner that identifies $\widehat \sigma$ with $\sigma$, as these both map between the two points covering a given point in $\widetilde \Sigma$. As $\widehat \sigma$ and $\sigma$ define the corresponding $\mathbb Z_2$ actions, this establishes that $\widehat M \cong (\mathbb R \times \widehat \Sigma)/ \mathbb Z_2 \cong (\mathbb R \times \Sigma)/ \mathbb Z_2 \cong M $.
\\ \qed

This theorem establishes that every possible orientable reflective topological big bang model may be uniquely and explicitly generated by considering the connected nonorientable manifolds $\widetilde \Sigma$, and the spatial topology $\Sigma$ of the corresponding expanding universe in each case is the orientation double cover of $\widetilde \Sigma$. Examples compatible with the possible homogeneous spatial geometries (constant positive, negative, or zero sectional curvature) may be constructed by seeking appropriate nonorientable quotients of $S^{n-1}$, $H^{n-1}$, and $\mathbb R^{n-1}$. It is interesting to note that in the case $n = 4$ of greatest interest, and more generally for any even $n$, Synge's theorem (Corollary 3.10 in Section 9 of \cite{do1992riemannian}) prohibits the existence of closed nonorientable $(n-1)$-manifolds with positive curvature, so there are no complete, oriented, constant positive curvature reflective topological big bang models (i.e.\@ no working quotients of $S^3$).

\begin{figure}[t]
\centering
\def\arraystretch{1.3}
\begin{tabular}{| c | c | c |} 
 \hline
Covering Chain & $\Sigma$ & $\widetilde \Sigma$ \\ [0.5ex] 
\hline
$\mathbb R^3 \to T^3 \to K^2 \times S^1$	    & $T^3$    & $K^2 \times S^1$
\\ \hline
$\mathbb R^3 \to T^2 \times \mathbb R \to K^2 \times \mathbb R$	    & $T^2 \times \mathbb R$    & $K^2 \times \mathbb R$
\\ \hline
$\mathbb R^3 \to T^2 \times \mathbb R \to M_s \times S^1$	    & $T^2 \times \mathbb R$    & $M_s \times S^1$
\\ \hline
$\mathbb R^3 \to S^1 \times \mathbb R^2 \to M_s \times \mathbb R$	    & $S^1 \times \mathbb R^2$    & $M_s \times \mathbb R$
\\ \hline
\end{tabular}
\caption{\small A few simple chains of covering relations giving admissible topologies of orientable reflective topological big bangs admitting flat metrics in $n=4$. Here, $K^2$ refers to the Klein bottle and $M_s$ the (open) M\"obius band.}
\label{fig:chaintable}
\end{figure}

Working topologies in $n = 4$ abound, however, in the zero curvature case, generally preferred by the standard $\Lambda$CDM model of cosmology \cite{planck}. 
A nonexhaustive list of a few simple topologies compatible with an orientable topological big bang model in $n = 4$ is shown in the table of Figure \ref{fig:chaintable}. 
One can see that there are examples compatible with both finite (e.g.\@ $\Sigma = T^3$) and infinite (e.g.\@ $\Sigma = T^2 \times \mathbb R$ or $\Sigma = S^1 \times \mathbb R^2$) spatial universes. 
Of course, these constant-curvature spaces are well-studied: a complete characterization of the flat quotients of $\mathbb R^3$, some of which are nonorientable, may be found in \cite{conway2003describing} (more comprehensively, see \cite{wolf2011spaces}). 
Working quotients of $H^3$ also exist in some abundance \cite{matsuzaki1998hyperbolic} (see also \cite{krasnov2000holography,krasnov2002analytic,krasnov2003holomorphic,manin2001holography} for discussion of $H^3$ quotients in the physical context of AdS/CFT holography in 2+1 dimensional gravity\footnote{These works extensively characterize quotients of 2+1 dimensional AdS (the most well-known being the BTZ black hole), finding a wealth of interesting structures. These may be viewed as analogous, in a sense, to Schr\"odinger's elliptic de Sitter space since the latter is a quotient of 3+1 dimensional de Sitter space. It is plausible that similarly exploring more general quotients in the present context may yield interesting pictures of cosmology.}), but we do not list or emphasize these here both since their construction is not nearly as simple as the examples shown and since standard cosmology typically prefers flat slicing.

\subsection{The Horizon Problem}
\label{sec:horizon}

So far, we have treated only the topological structure of a reflective topological big bang model. To showcase such a model's geometric structure, we now discuss its interplay with cosmology's horizon problem. This is a point on which a reflective model distinctly deviates from the related bouncing model.

We consider a spatially flat FLRW metric, given locally by
\begin{equation} 
ds^2 = -dt^2 + a(t)^2 (dx^2 + dy^2 + dz^2), \label{eqn:flrw}
\end{equation}
for now on $\mathbb R_+ \times \Sigma$ (for any $\Sigma$ in Figure \ref{fig:chaintable}, say) as in standard cosmology. The scale factor $a: \mathbb R_+ \to \mathbb R_+$ is coupled to matter through the Friedmann equations,
\begin{align} 
H^2 & = \frac{8 \pi}{3} \rho, \label{eqn:fried1}
\\ \frac{\ddot a}{a} & = - \frac{4 \pi}{3} (\rho + 3 P), \label{eqn:fried2}
\end{align}
with $H(t) := \frac{\dot a}{a}$ the Hubble parameter and the energy densities $\rho = \sum_i \rho_i$ and pressures $P = \sum_i P_i$, all functions of $t$, summed across the matter components. These imply the closely related fluid equation,
\begin{equation}
\dot \rho = -3H(\rho + P). \label{eqn:fluid}
\end{equation}
Neglecting energy transfer between components, each satisfies equation (\ref{eqn:fluid}) individually with dynamics entirely characterized by its constant equation of state $w_i := P_i/\rho_i$. Standard (post-inflationary) cosmology incorporates:
\begin{itemize}
\item Baryonic and dark matter $\rho_m$ with $w_m = 0$.
\item Radiation $\rho_\gamma$ with $w_\gamma = 1/3$.
\item Dark energy $\rho_\Lambda$ with $w_\Lambda = -1$.
\end{itemize} 
Equation (\ref{eqn:fluid}) implies that each component satisfies $\rho_i \propto a^{-3(1+w_i)}$, so we may rewrite equation (\ref{eqn:fried1}) in terms of the present ($a=1$) Hubble parameter $H_0$ as
\begin{equation} \label{eqn:hubble}
H^2 = H_0^2 \left[ \Omega_\Lambda + \frac{\Omega_m}{a^3} + \frac{\Omega_\gamma}{a^4} \right].
\end{equation}
In Planck's best-fit of $\Lambda$CDM cosmology to a variety of observables \cite{planck}, $\Omega_m \approx 0.3144$, $\Omega_{\Lambda} \approx 0.6855$, and $\Omega_\gamma \approx 9.23 \cdot 10^{-5}$.

The {\it horizon problem} of standard cosmology is the observation that the co-moving distance $d_{CMB}$ traversed by signals between recombination at $a_{rec} \approx 1/1100$ and the present,
$$ d_{CMB} = \int_{t_{rec}}^{t_{now}} \frac{dt}{a(t)} = \int_{a_{rec}}^1 \frac{da}{a^2 H} \sim \frac{2}{H_0 \sqrt{\Omega_m}},$$
is much greater than the co-moving distance $d_{causal}$ traversed by signals between the big bang and recombination,

$$ d_{causal} = \int_{0}^{t_{rec}} \frac{dt}{a(t)} = \int_{0}^{a_{rec}} \frac{da}{a^2 H} \sim \frac{2 \sqrt{a_{rec}}}{H_0 \sqrt{\Omega_m}} \approx d_{CMB}/33,$$
indicating that patches of the CMB seen at antipodal points in the sky were not in causal contact at the time of emission. Standard cosmology resolves this by introducing an inflationary era in the early universe, modifying the early dynamics of $a(t)$ to arrange that $d_{causal}/d_{CMB} \gtrapprox 1$.

What might change in this picture in a reflective topological big bang? The most important point is that the spacetime manifold still perfectly admits the spatially homogeneous metric (\ref{eqn:flrw}), with the only additional requirements for $C^2$ smoothness across the reflection surface being $\dot a(0) = 0$ and $a_0 := a(0) > 0$. 
According to equation (\ref{eqn:fried1}), the condition $\dot a(0) = 0$ requires that $\rho \to 0$ at the reflection surface. 
As we are working classically, we do not attempt to consider how this might occur in the complicated quantum-laden physics near the big bang for $a_0$ sufficiently small, but in a collection of classical perfect fluids, it would require at least one fluid to contribute with a negative energy density\footnote{Bouncing cosmologies must also wrangle with the requirement $\dot a(0) = 0$, and works in that direction have considered better-motivated matter models (see Section III.C of \cite{battefeld2015critical}), generally requiring a violation of the null energy condition near $t = 0$. We invoke this simplest possible implementation to exhibit basic features and make a simple computation.} of magnitude comparable to the dominant positive component near $a_0$.

To obtain a minimal classical model of a $C^2$ reflective topological big bang which agrees with standard cosmology at late times, then, we add to our mix of matter components an ``exotic" fluid satisfying $\rho_{ex} < 0$ with $w_{ex} > 1/3$. Taking $p : = 3(1+w_{ex}) > 4$, equation (\ref{eqn:hubble}) now becomes

\begin{equation} \label{eqn:hubbletr}
H^2 = H_0^2 \left[ \Omega_\Lambda + \frac{\Omega_m}{a^3} + \frac{\Omega_\gamma}{a^4} - \frac{\Omega_{ex}}{a^p} \right],
\end{equation}
with $\Omega_{ex} << \Omega_\gamma$ to ensure agreement with standard cosmology except at the earliest times. Of course, we do not at all claim that equation (\ref{eqn:hubbletr}) reflects a realistic model of the physical universe, but only that it is a minimal adjustment to standard cosmology yielding a geometry compatible with a reflective topological big bang, allowing simple analysis of qualitative features. That $p > 4$ ensures that there is an earliest scale factor $a_0$ at which $\dot a = a H = 0$. As this should occur long before matter-radiation equality, we neglect $\Omega_m$ and $\Omega_\Lambda$ to write
$$ a_0 \approx \left( \frac{\Omega_{ex}}{\Omega_\gamma} \right)^{1/(p-4)}.$$

While equation (\ref{eqn:fried2}) indicates that $\ddot a > 0$ at $a_0$, this switches sign at a scale factor $a_1$ given by
$$ a_1 \approx \left( \frac{3w_{ex}+1}{2} \frac{\Omega_{ex}}{\Omega_\gamma} \right)^{1/(p-4)} = \left( \frac{2+\epsilon}{2} \right)^{1/\epsilon} a_0,$$
where we've defined $\epsilon$ by $p = 4 + \epsilon$. Assuming $a_1 << a_{rec}$ so as not to interfere with standard cosmology, we now constrain $d_{causal}$:

$$ d_{causal} = \int_{0}^{t_{rec}} \frac{dt}{a(t)} >  \int_{0}^{t_1} \frac{dt}{a(t)} = \int_{a_0}^{a_1} \frac{da}{a^2 H} \approx  \frac{1}{H_0 \sqrt{\Omega_\gamma}} \int_{a_0}^{a_1} \frac{da}{\sqrt{1 - \frac{\Omega_{ex}}{\Omega_\gamma} a^{-\epsilon}}}. $$
Changing variables to $u := \frac{\Omega_{ex}}{\Omega_\gamma} a^{-\epsilon}$, i.e. $a = \left( \frac{1}{u} \frac{\Omega_{ex}}{\Omega_\gamma} \right)^{1/\epsilon} = u^{-1/\epsilon} a_0$, we find

\begin{align*} 
\int_{0}^{t_1} \frac{dt}{a(t)} 
& \approx \frac{a_0}{H_0 \sqrt{\Omega_{\gamma}} \epsilon} \int_{\frac{2}{2+\epsilon}}^1 \frac{ u^{-\frac{1}{\epsilon}} du}{u \sqrt{1 - u }} 
\\ & > \frac{a_0}{H_0 \sqrt{\Omega_{\gamma}} \epsilon} \int_{\frac{2}{2+\epsilon}}^1 \frac{du}{u \sqrt{1 - u }} 
= \frac{a_0}{H_0 \sqrt{\Omega_{\gamma}}} \cdot \frac{2}{\epsilon} \tanh^{-1} \left( \sqrt{\frac{\epsilon}{2+\epsilon}} \right) .
\end{align*}
In particular, as $\epsilon \to 0$ we have
$$ d_{causal} \gtrapprox \frac{a_0}{H_0 \sqrt{\Omega_\gamma}} \sqrt{\frac{2}{\epsilon}} \approx d_{CMB} \cdot \sqrt{\frac{ a_0^2 \Omega_m}{2 \Omega_\gamma \epsilon} } = d_{CMB} \cdot \frac{ a_0}{\sqrt{2 a_{eq} \epsilon} },$$
where $a_{eq} = \frac{\Omega_\gamma}{\Omega_m} \approx 2.94 \cdot 10^{-4}$ is the scale factor of matter-radiation equality. It may appear, then, that $d_{causal}/d_{CMB} \to \infty$ like $1/\sqrt{\epsilon}$ as $\epsilon \to 0$, resolving the horizon problem (without inflation) for $\epsilon$ sufficiently small, but this neglects the dependence on $a_0$. Indeed, resolving the horizon problem via this estimate would require
$$ \epsilon \lessapprox \frac{a_0^2}{2a_{eq}},$$
but the scale factor $a_2$ at which the exotic fluid becomes negligible, e.g.\@ such that $\frac{\Omega_\gamma}{a_2^4} = 10 \cdot \frac{\Omega_{ex}}{a_2^p}$, would then satisfy (since $10^{1/\epsilon}\sqrt{\epsilon} \geq \sqrt{2e \ln(10)}$ for all $\epsilon > 0$)
$$a_2 = 10^{1/\epsilon} a_0 \geq \sqrt{2 a_{eq} \epsilon} \cdot 10^{1/\epsilon}  \geq 2\sqrt{e \ln(10) a_{eq}} \approx 0.086 >> a_{eq},$$
strongly modifying standard cosmology in well-tested regimes.

In the other direction, if we directly impose $a_{eq} > a_2 = 10^{1/\epsilon} a_0$, then the modification to the quantity $d_{causal}/d_{CMB}$ from standard cosmology (essentially entirely occurring between $a_0$ and $a_2$) satisfies\footnote{The supremum in this inequality chain is approximated numerically.}

\begin{align*} 
\frac{\Delta d_{causal}}{d_{CMB}} 
& \approx \frac{1}{d_{CMB}} \int_{0}^{t_2} \frac{dt}{a(t)} 
\approx \frac{a_0}{2\sqrt{a_{eq}} \epsilon} \int_{1/10}^1 \frac{ u^{-\frac{1}{\epsilon}} du}{u \sqrt{1-u}} 
= \frac{a_2}{2\sqrt{a_{eq}} \epsilon} \int_{1/10}^1 \frac{ (10u)^{-\frac{1}{\epsilon}} du}{u \sqrt{1-u}} 
\\ & \leq \frac{a_2}{2 \sqrt{a_{eq}}} \cdot  \sup_{s > 0} \left(  s \int_{1/10}^1 \frac{ (10u)^{-s} du}{u \sqrt{1-u}} \right)
\approx  0.571 \cdot \frac{a_2}{\sqrt{a_{eq}}}
\\ & < 0.571 \cdot \sqrt{a_{eq}} \approx 0.01,
\end{align*}
meaning that the horizon problem cannot be resolved (at least in the simple implementation of equation (\ref{eqn:hubbletr})) without taking $a_2 > a_{eq}$ and significantly impacting standard cosmology. This suggests that a reflective topological big bang cosmology still requires a process such as inflation in the early universe to resolve the horizon problem in a manner consistent with observations.

\section{Beyond Reflection}
\label{sec:beyond}

Throughout Section \ref{sec:reflective}, we have taken some care to emphasize the qualifier {\it reflective} in our discussion of topological big bangs. This is because one might imagine that it is possible to replace the singularity of cosmological models with some topology in a manner qualitatively different than Definition \ref{def:topbigbang}, but still consistent with the broader idea of a topological big bang. We exhibit that this is indeed the case with an explicit example.

\subsection{The Itty-Bitty Blender} \label{subsec:ittybitty}
\begin{figure}[t]
\centering
\includegraphics[width=0.55\textwidth]{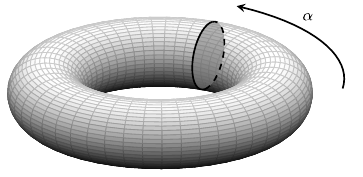}
\hspace{0.5cm}
\begin{tikzpicture}
\filldraw(0,0) circle (1.5pt);
\def\r{1.75}
\def\a{0.8}
\filldraw({-\a*\r/sqrt(2)},{\a*\r/sqrt(2)}) circle (2pt);
\filldraw[opacity=0.25] (0,0) circle (\r);
\draw[thick] (0,0) -- node[below left]{$r$} ({-\a*\r/sqrt(2)},{\a*\r/sqrt(2)});
\draw[dashed] (0,0) -- (\r,0);
\draw ({\a*\r/2},0) arc[radius=({\a*\r/2}), start angle=0, delta angle={135}] node[midway,above]{$\theta$};
\end{tikzpicture}
\caption{\small Coordinates on the solid toroidal region $r<1$. $\alpha$ specifies the shaded cross section (left), and $(r,\theta)$ are polar coordinates within the cross section.}
\label{fig:coordinates}
\end{figure}
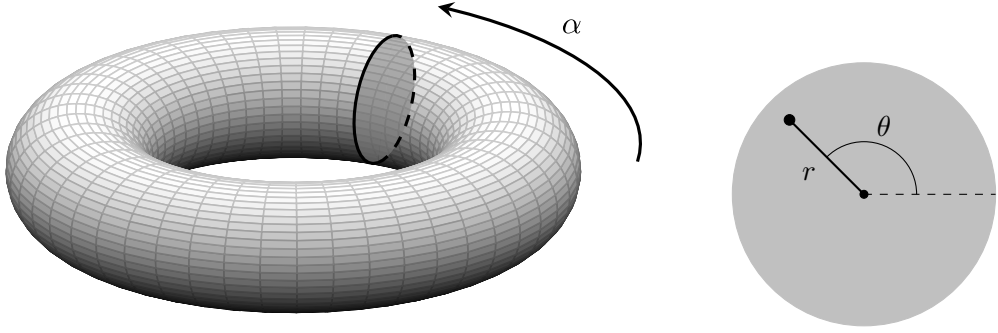

We consider the smooth manifold $M := \mathbb R^2 \times T^2 = (\mathbb R^2 \times S^1) \times S^1$. Letting $\alpha,\beta$ be the two angular coordinates on the $S^1$ copies and $(r,\theta)$ be polar coordinates on $\mathbb R^2$ (so each of $\theta,\alpha, \beta$ run over $[0,2\pi)$), consider the metric
\begin{equation} \label{eqn:blendermetric}
g = (1-r^2) dr^2 - 4r dr d\alpha + (r^2-1) d\alpha^2 + r^2 d \theta^2 + (r^2+1) d \beta^2
\end{equation}
This is regular across $r = 0$ since the $\mathbb R^2$ piece may be expressed in Cartesian coordinates $(x,y)$ as 
\begin{equation}
(1-r^2) dr^2 + r^2 d \theta^2 = dx^2 + dy^2 - (x dx+ y dy)^2.
\end{equation}
Note that the coordinate vector field $\partial_\beta$ is always spacelike, so the last $S^1$ factor does not interestingly affect the overall structure; we suppress this factor in visualizations and discussion. 
In contrast, $\partial_r$ and $\partial_\alpha$ exchange causal character at $r = 1$. 
For $r < 1$, $\partial_r$ is spacelike and $\partial_\alpha$ is timelike: in this range, one should envision that $(r,\theta,\alpha)$ parameterizes a solid toroidal region with $r$ the radial coordinate in the circular cross section at azimuthal angle $\alpha$, and with $\theta$ the angular coordinate around this cross section (see Figure \ref{fig:coordinates}). 

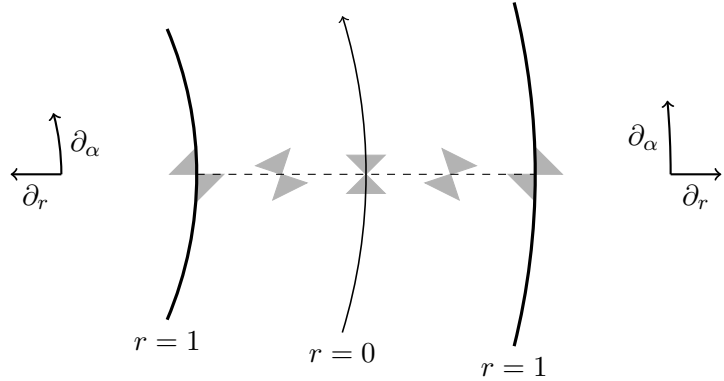
\begin{figure}[t]
\centering
\begin{tikzpicture}[scale=0.9]
\def\R{8} \def\r{2.5}
\def\lang{23} \def\mang{17} \def\rang{14} \def\sang{5}
\def\opa{0.3}
\def\tl{0.4}
\draw[very thick] ({-\R+(\R-\r)*cos(\lang)},{-(\R-\r)*sin(\lang)}) node[below]{$r=1$} arc[radius=({\R-\r}), start angle={-\lang}, delta angle={2*\lang}];

\draw[very thick] ({-\R+(\R+\r)*cos(\rang)},{-(\R+\r)*sin(\rang)}) node[below]{$r=1$} arc[radius=({\R+\r}), start angle={-\rang}, delta angle={2*\rang}];

\draw[->,semithick] ({-\R+\R*cos(\mang)},{-\R*sin(\mang)}) node[below]{$r=0$} arc[radius=({\R}), start angle={-\mang}, delta angle={2*\mang}];

\filldraw[opacity=\opa] ({-\tl/sqrt(2)},{-\tl/sqrt(2)}) -- (0,0) -- ({\tl/sqrt(2)},{-\tl/sqrt(2)}) -- cycle;
\filldraw[opacity=\opa] ({-\tl/sqrt(2)},{\tl/sqrt(2)}) -- (0,0) -- ({\tl/sqrt(2)},{\tl/sqrt(2)}) -- cycle;

\filldraw[opacity=\opa] ({\r/2-3*\tl/sqrt(10)},{-\tl/sqrt(10)}) -- (\r/2,0) -- ({\r/2+\tl/sqrt(10)},{-3*\tl/sqrt(10)}) -- cycle;
\filldraw[opacity=\opa] ({\r/2-\tl/sqrt(10)},{3*\tl/sqrt(10)}) -- (\r/2,0) -- ({\r/2+3*\tl/sqrt(10)},{\tl/sqrt(10)}) -- cycle;

\filldraw[opacity=\opa] ({-\r/2+3*\tl/sqrt(10)},{-\tl/sqrt(10)}) -- (-\r/2,0) -- ({-\r/2-\tl/sqrt(10)},{-3*\tl/sqrt(10)}) -- cycle;
\filldraw[opacity=\opa] ({-\r/2+\tl/sqrt(10)},{3*\tl/sqrt(10)}) -- (-\r/2,0) -- ({-\r/2-3*\tl/sqrt(10)},{\tl/sqrt(10)}) -- cycle;

\filldraw[opacity=\opa] (\r-\tl,0) -- (\r,0) -- (\r,-\tl) -- cycle;
\filldraw[opacity=\opa] (\r,\tl) -- (\r,0) -- (\r+\tl,0) -- cycle;

\filldraw[opacity=\opa] (-\r-\tl,0) -- (-\r,0) -- (-\r,\tl) -- cycle;
\filldraw[opacity=\opa] (-\r,-\tl) -- (-\r,0) -- (-\r+\tl,0) -- cycle;

\draw[->,thick] (\r+2,0) arc[radius=({\R+\r+2}), start angle={0}, delta angle={\sang}] node[midway,left]{$\partial_\alpha$};
\draw[->,thick] (\r+2,0) -- node[midway,below]{$\partial_r$} (\r+2.75,0);

\draw[->,thick] (-\r-2,0) arc[radius=({\R-\r-2}), start angle={0}, delta angle={3*\sang}] node[midway,right]{$\partial_\alpha$};
\draw[->,thick] (-\r-2,0) -- node[midway,below]{$\partial_r$} (-\r-2.75,0);

\draw[dashed] (-\r,0) -- (\r,0);
\end{tikzpicture}
\caption{\small A top-down view of a section of the torus, showing rotation of timecones on the region $r < 1$ as $r$ varies, according to equation (\ref{eqn:nulls}). If one continued to $r > 1$, the timecones would asymptote to rotating a full 90 degrees outward.}
\label{fig:timecones}
\end{figure}

As $\partial_\alpha$ is timelike here, each ring around the torus at fixed $r < 1$ and $\theta$ is a closed timelike curve, so matter in this region could repeatedly swirl around the torus: this is the ``blender". As one increases $r$ from the central ring, however, the timecones rotate from pointing parallel to the azimuthal direction at $r = 0$ to pointing more and more outwards at larger $r$ (see Figure \ref{fig:timecones}). One can both track this rotation and establish that the spacetime is indeed time-orientable by noting that the smooth vector field $T := (\partial_\alpha + r \partial_r)/(1+r^2)$ is unit timelike, as
\begin{equation}
\langle \partial_\alpha + r \partial_r, \partial_\alpha + r \partial_r \rangle = r^2(1-r^2) - 4r^2 + r^2-1 = -r^4-2r^2-1 = -(1+r^2)^2.
\end{equation}
We declare $T$ to be future-directed. Similarly, one can verify that two future-directed null directions $N_\pm$ satisfying $\langle N_+, N_- \rangle = -2$ and $2T = N_+ + N_-$ are given by
\begin{equation} \label{eqn:nulls}
(1+r^2) N_{\pm} = (\partial_\alpha + r \partial_r) \pm (\partial_r - r \partial_\alpha) = (r \pm 1) \partial_r + (1\mp r)\partial_\alpha.
\end{equation}

If one thickens the torus beyond $r = 1$, then the timecones are asymptotically parallel to the radial direction at large $r$: the null directions $N_{\pm}$ are asymptotically proportional to $\partial_r \mp \partial_\alpha$, symmetric about $T \to \partial_r/r$, the radial direction. This means that typical future-directed timelike curves swirl around the blender many times, but they leave permanently toward $r \to \infty$ if they ever venture too far outward (to $r > 1$).

So, what does this spacetime have to do with cosmology? First note that, purely topologically, if one excises the central $r = 0$ ring from the blender, then $M \backslash \{r = 0\} \cong \mathbb R_+ \times T^3$, with $\mathbb R_+$ parameterized by $r$ and $T^3$ by $(\theta,\alpha,\beta)$. As discussed above, we may naturally think of $r$ as a time coordinate for $r >> 1$, making the $T^3$ factors (finite) spatial slices, and the metric (\ref{eqn:blendermetric}) asymptotes to
\begin{equation}
g \approx r^2 \left[ -dr^2 + d \theta^2 + d \alpha^2 + d \beta^2 \right],
\end{equation}
a spatially flat FLRW metric expressed in conformal time. Changing coordinates to the apparent comoving time $t := r^2/2$ turns equation (\ref{eqn:blendermetric}) into
\begin{equation} \label{eqn:blendermetric_t}
g = -\left( 1-\frac{1}{2t} \right) dt^2 - 4 dt d\alpha + (2t-1) d\alpha^2 + 2t d \theta^2 + (2t+1) d \beta^2.
\end{equation}
This metric asymptotes, for $t >> 1$, to the standard FLRW metric (\ref{eqn:flrw}) with apparent scale factor $a(t) = \sqrt{2t}$, corresponding to a flat, {\it radiation-dominated} universe! Somewhat surprisingly upon first glance, it smoothly extends to $t = 0$ by virtue of our construction. 
The blender, with some rescaling, can then simply append to the standard cosmology's very early universe in the radiation-dominated regime, supplanting the singularity: it is ``itty-bitty" in that we think to append it when the (finite) universe is very young and highly contracted.

As in the case of reflective topological big bangs, the causal pathologies of this spacetime (closed timelike curves) are localized to the very early universe. One may still make sense of a classical initial value problem for the future of any spatial $T^3$ slice outside the blender (at any $r_0 > 1$), and crucially, an observer in the present has no way of accessing a blender cosmology's closed timelike curves. It is interesting to note that there is no need for inflation to resolve the horizon problem in this picture: the eternal mixing accommodated by the blender ensures that the whole of spacetime is thoroughly causally embroiled, as every inextendible causal curve emanated from the blender. This resolution is fundamentally distinct from that of inflation, painting a rather different picture of how the early universe may have thermally equilibrated. As pathological as dynamically emerging closed timelike curves would be \cite{hawking1992chronology}, their existence at the ``beginning" of the universe in this manner, perhaps in the regime of quantum gravity and supported by some sort of self-consistent steady-state ``swirling" matter distribution, may be permissible.

As discussed in the introduction, we do not attempt here to replicate other successes of inflationary perturbative cosmology (see \cite{achucarro2022inflation}) or claim that the present model is a genuinely compelling alternative on its own without much more work. Indeed, the metric (\ref{eqn:blendermetric}) is somewhat contrived and was composed with no thought toward physical energy conditions-- in fact, a computation shows that it satisfies none of the standard energy conditions in the $r < 1$ region. It is an interesting question whether a qualitatively similar picture can be attained under, say, the dominant energy condition or via a physically motivated matter field; this is discussed further below. Instead, this example, as the whole of this work, is a proof of concept that there exists a rich vein of possibilities for the nature of the early universe beyond the standard discourse.

\subsection{The Blender Unwound: An Eternal Trumpet} \label{subsec:trumpet}
In both reflective topological big bangs and the Itty-Bitty Blender spacetime, some sort of causal pathology is introduced in the early moments of the model that supplant the singularity of standard cosmology. 
One might get the impression that causal pathologies are a {\it requirement} for smoothing out the singularity, particularly if one is to respect energy conditions. 
The singularity theorems of Hawking and Penrose \cite{penrosesingularity,hawking1966occurrence,hawking1966occurrence2,hawking1967occurrence,hawking1970singularities} (see \cite{senovilla20151965} for a modern review), one iteration of which is recalled below from O'Neil's text (Theorem 55A in \cite{oneil}), indeed say something to this effect:

\begin{theorem} \label{thm:hawking1}
    {\bf (Hawking, Version 1)} Suppose $\text{Ric}(v,v) \geq 0$ for every timelike tangent vector $v \in T_pM$. Let $S \subset M$ be a spacelike future Cauchy hypersurface with future convergence $k \geq b > 0$. Then every future-pointing timelike curve through $M$ starting in $S$ has length at most $1/b$.
\end{theorem}

Supposing that one accepts the curvature hypothesis, known as the {\it strong energy condition}, this is applied in time-reversed form to cosmology by identifying $S$ as a spatial slice $\Sigma_t = \{t\} \times \Sigma$, on which the convergence is measured by the (notably nonzero) Hubble parameter $H(t)$. 
While this theorem makes a remarkably strong conclusion, the curvature hypothesis is generally seen as too strong: it is not satisfied by a universe dominated, at some point in the past of $\Sigma_t$, by a cosmological constant (or, say, a scalar field masquerading as one). 
In this way, inflation models are unconstrained by this particular theorem, seemingly allowing them to avoid a singularity. We note, however, that more modern variants of such theorems \cite{borde1996singularities,borde2003inflationary,guth2007eternal} indicate that typical inflation models should still be past-incomplete. 

Reflective topological big bangs and the Itty-Bitty Blender have an additional out through their causal pathologies-- they are clearly not globally hyperbolic, so the cosmological spatial slices $\Sigma_t$ are not Cauchy surfaces. 
While it is clear from their statements, and therefore widely recognized \cite{senovilla20151965}, that sufficiently egregious causal pathologies allow one to work around most singularity theorems, the present examples are interesting in that they illustrate that such pathologies can be entirely localized to the universe's ``earliest" moments.
It may appear conceivable, then, that one could write down a metric respecting the strong energy condition while having the same qualitative structure as the Itty-Bitty Blender. 
This is not the case for the blender as we have presented it, with cosmological spatial slices being copies of $T^3$ (and hence compact), due to the following variant of Hawking's theorem (Theorem 55B in \cite{oneil}; a relevant result is also established in \cite{maeda1996causality}):

\begin{theorem} \label{thm:hawking2}
    {\bf (Hawking, Version 2)} Suppose $\text{Ric}(v,v) \geq 0$ for every timelike tangent vector $v \in T_pM$. Let $S \subset M$ be a compact spacelike hypersurface with future convergence $k > 0$. Then $M$ is future timelike incomplete.
\end{theorem}
Any blender-type construction on $\mathbb R^2 \times T^2$, then, cannot respect the strong energy condition without introducing timelike incompleteness. 
One could reasonably accept this and seek a metric only satisfying an independent (and more physical) energy condition, such as the dominant energy condition\footnote{Recall that the dominant energy condition imposes that any four-momentum density seen by observers is future-pointing and causal, a directly physical requirement. In contrast, while the strong energy condition is geometrically convenient, it can be violated by entirely reasonable and simple classical matter models, such as free scalar fields.}-- there appears to be no obstruction to this in the literature:

\begin{question} \label{quest:modify-blender}
Does there exist a geodesically complete and time-orientable Lorentzian metric on $\mathbb R^2 \times T^2$ which respects the dominant energy condition and asymptotes to an FLRW geometry?
\end{question}

It would be very interesting for considerations of what constitutes a viable model of cosmology if the answer to this question were affirmative. 
Another approach to working around the constraints of Theorems \ref{thm:hawking1} and \ref{thm:hawking2} is to ``unwind" one or both factors of $S^1$ to factors of $\mathbb R$, giving noncompact cosmological spatial slices with topology either $T^2 \times \mathbb R$ or $S^1 \times \mathbb R^2$. It is important to note that each of equations (\ref{eqn:blendermetric}) through (\ref{eqn:blendermetric_t}) lift precisely as stated to these ``unwound" spacetimes, so some essential features are preserved: namely, the metric is still asymptotic to that of a radiation-dominated FLRW spacetime.

\begin{figure}[b!]
\centering
\includegraphics[width=0.5\textwidth]{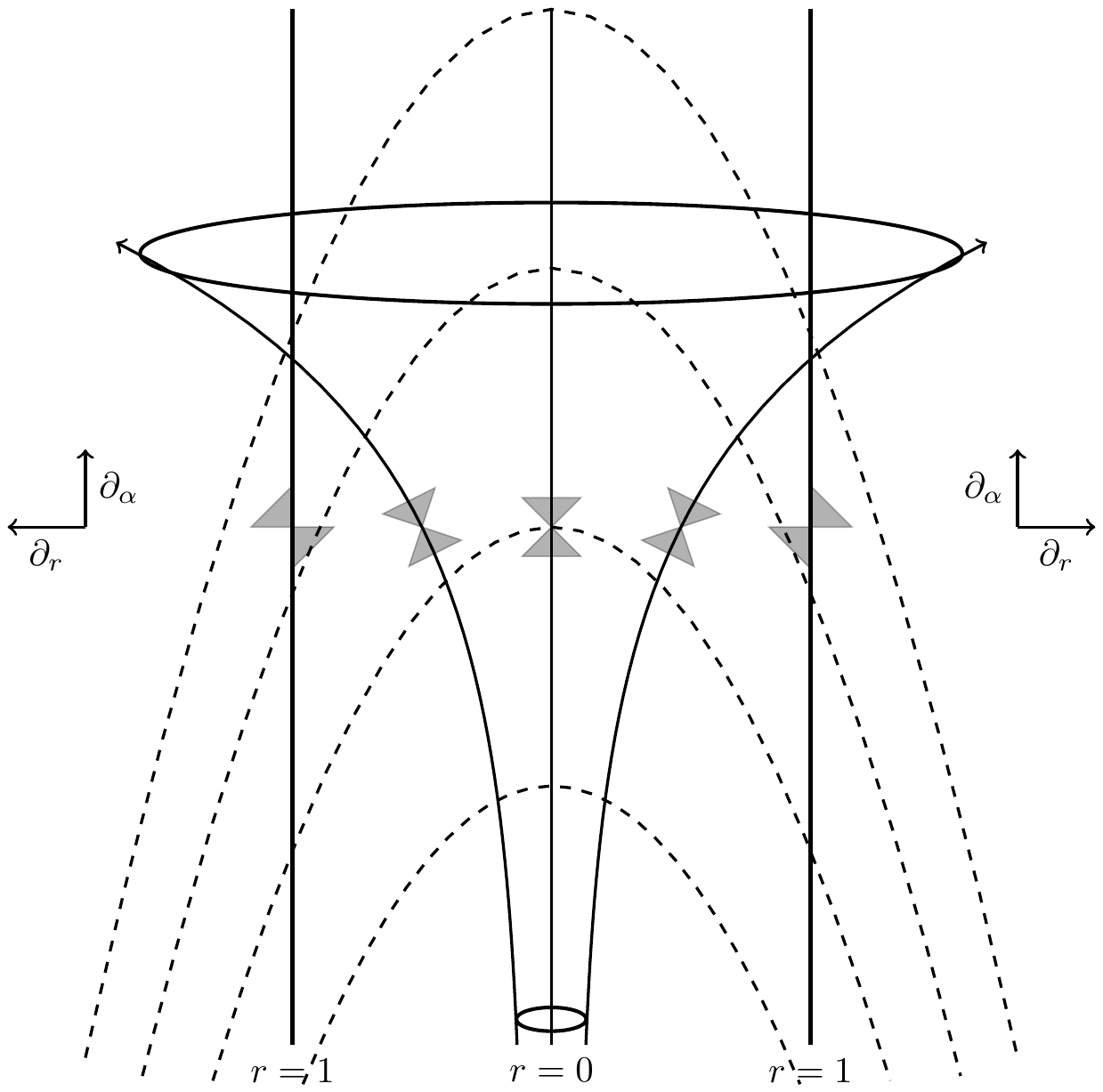}
\caption{\small A cross-section of the tubular core of the Eternal Trumpet spacetime. Dashed curves are cross-sections of the foliating paraboloid Cauchy surfaces. The two diverging curves resembling the ``bell" of a trumpet are representative future-directed timelike curves exiting the core toward the cosmological region $r>>1$. The full spacetime is obtained by rotating about the $r = 0$ axis (illustrated via added ellipses on the trumpet) and crossing with $\mathbb R$.}
\label{fig:trumpet}
\end{figure}

An especially interesting example, dubbed the ``Eternal Trumpet", emerges upon passing to the universal cover of the blender by unwinding both factors, so that the full spacetime has the trivial topology of $\mathbb R^4$. Suppressing the $\beta$ factor as before,
the solid toroidal core of the blender is now unwound into an infinite solid cylindrical tube, which means that the curves in the illustration of Figure \ref{fig:timecones} have simply straightened out and extend vertically indefinitely. 
In particular, what were closed timelike curves proceeding around the blender's toroidal core now proceed along the tubular core indefinitely. A generic future-directed timelike curve proceeds along the tubular core for some time before eventually leaving to the $r > 1$ region and necessarily turning radially outwards, limiting to $r \to \infty$.
A congruence of such curves all exiting the core at the same value of $\alpha$ bend outwards and away from the core in all directions, resembling the shape of a trumpet eternally ``blowing" these trajectories out of the core and into the cosmological region (see Figure \ref{fig:trumpet}).
The effective cosmological slices are topologically $S^1 \times \mathbb R^2$, cylindrical shells at fixed $r >> 1$. 
This picture of cosmology also has no horizon problem, as all points share a causal past in the tubular core.

Perhaps the most interesting feature of this unwound picture is that the causal improvement over the blender is much stronger than simply eliminating the closed timelike curves: the spacetime is now globally hyperbolic, foliated by paraboloid Cauchy surfaces (see Figure \ref{fig:trumpet}) that are level sets of the time function 
$$f(r,\theta,\alpha,\beta) = \alpha+r^2/2.$$ 
Indeed, the unit timelike vector field $T = (\partial_\alpha + r \partial_r)/(1+r^2)$ satisfies $T = - \nabla f$. 
This spacetime is apparently as causally regular as one could hope, asymptotes to a radiation-dominated FLRW geometry, and is geodesically complete. 
While, as mentioned above, the given metric (\ref{eqn:blendermetric}) satisfies none of the standard energy conditions in the $r < 1$ region, a notable observation is that this is not forced by either Theorem \ref{thm:hawking1} or Theorem \ref{thm:hawking2}, as the noncompact effective cosmological slices at fixed $r >> 1$ are not the Cauchy surfaces of this globally hyperbolic spacetime (their past Cauchy horizons are given by the $r = 1$ surface of the tubular core).
This illustrates that when considering variants of the singularity theorems formulated for globally hyperbolic models of cosmology, then, one must keep in mind that even if the spacetime is globally hyperbolic, the cosmological slices empirically expected to exhibit geodesic convergence may not be among its Cauchy surfaces. In the present example, the foliating paraboloids are ``almost" the cosmological slices for very large $r >> 1$, but bend away from them near $r \approx 1$.

From the perspective of an initial value problem, an alternative phrasing of these observations is that the $r >> 1$ cosmological slices illustrate the existence of initial data sets that are arbitrarily small perturbations of an FLRW initial data set taken in the radiation-dominated regime and that admit a past Cauchy development that is indeed timelike incomplete, as expected from Theorem \ref{thm:hawking1}, but which is smoothly extendable to a geodesically complete and globally hyperbolic spacetime. That is, one can apply Theorem \ref{thm:hawking1} to one of the effective cosmological slices in the Eternal Trumpet spacetime, but the conclusion of incompleteness only applies to its Cauchy development (more precisely, to that portion of the development respecting the strong energy condition), and that development is smoothly extendable to a spacetime which is complete.
Due to the strong curvature nature of the initial singularity in standard FLRW spacetimes, it is often thought that ``singularities" guaranteed by results such as Theorem \ref{thm:hawking1} in a perturbation should also be physical: The Eternal Trumpet demonstrates that this may not be the case. 

This discussion leads to the natural question of whether one can emulate the Eternal Trumpet's structure while respecting a physical energy condition, which would make the observations of the previous paragraph much more compelling:
\begin{question} \label{quest:modify-trumpet}
Does there exist a timelike geodesically complete and globally hyperbolic Lorentzian metric on $\mathbb R^4$ which respects the dominant energy condition and asymptotes to an FLRW geometry?
\end{question}

Note that, despite the discussion of the previous few paragraphs, we have not asked that the metric satisfy the strong energy condition; this is because one of Hawking's original results (Theorem 2 of \cite{hawking1967occurrence}) replaces the Cauchy surface hypothesis of Theorem \ref{thm:hawking1} with a more local integral condition that would likely be met by virtue of asymptoting to an FLRW geometry.
We have also not asked that it be fully geodesically complete; this is because, in light of global hyperbolicity, Penrose's result \cite{penrosesingularity} does not allow for null geodesic completeness under the null energy condition (implied by the dominant energy condition) if the FLRW region is spatially large enough that it includes trapped surfaces (e.g.\@ a sphere larger than the Hubble radius in some cosomological slice, which a model of our universe should likely admit).
To the best of our knowledge, however, it seems that one has hope of avoiding all known singularity theorems, even under the strong energy condition, if one only unwinds the $\beta$ factor of $S^1$ in the Itty-Bitty Blender spacetime, eliminating the compactness of the cosmological slices (rendering them $\mathbb R \times T^2$) but not the blender's localized closed timelike curves. 
This leaves one more interesting question:
\begin{question} \label{quest:modify-between}
Does there exist a geodesically complete and time-orientable Lorentzian metric on $\mathbb R^3 \times S^1$ which respects all energy conditions and asymptotes to an FLRW geometry?
\end{question}
An interesting observation is that, in principle, it seems there is no known obstruction to answering this affirmatively using only radiation as one's matter source. 
We remark that in attempting to construct a metric resolving any of Questions \ref{quest:modify-blender}, \ref{quest:modify-trumpet}, or \ref{quest:modify-between}, one must keep in mind the constraints imposed by unwinding and rewinding: a metric similar to (\ref{eqn:blendermetric}) and affirmatively resolving Question \ref{quest:modify-between}, for example, cannot be so symmetric that it can quotient to $\mathbb R^2 \times T^2$, nor can it be so nice that it lifts to a globally hyperbolic metric on $\mathbb R^4$.

\section{Conclusion} \label{sec:conclusion}

In this work, we have explored a variety of smooth and complete spacetime structures that may be appended to the FLRW background underlying modern $\Lambda$CDM cosmology in place of an initial singularity. The most minimal implementation, a reflective topological big bang, generalizes elliptic de Sitter space and involves allowing spacetime to smoothly wrap around on itself in ``reflecting" off of an earliest instant in time (more formally, see Definition \ref{def:topbigbang}). As reflective topological big bangs are not time-orientable near the reflection, this may be viewed as replacing the total pathology of the singularity with a lesser, localized pathology in causality. 

In Theorem \ref{thm:classification}, we have classified the admissible orientable (not time-orientable) topologies of reflective topological big bangs in $n$ spacetime dimensions as being in one-to-one correspondence with connected, nonorientable $(n-1)$-manifolds. For such manifolds which arise as quotients of $\mathbb R^{n-1}$ (see Figure \ref{fig:chaintable}), these topological big bang models may be endowed with a standard flat FLRW metric (\ref{eqn:flrw}), with only a minor constraint on the scale factor required for smoothness across the reflection. In Section \ref{sec:horizon}, we computed that minimal reflection models do not appear to be able to resolve the horizon problem on their own-- a mechanism such as inflation may still be required.

In Section \ref{subsec:ittybitty}, we considered a qualitatively different, nonreflective mollification of the singularity in the Itty-Bitty Blender spacetime. 
This was specified via an explicit metric (\ref{eqn:blendermetric}) placed on $\mathbb R^2 \times T^2$ which is asymptotic to that of a radiation-dominated FLRW geometry, so that the blender's nontrivial topology may be smoothly adjoined to the radiation-dominated era in standard cosmology. 
While time-orientable, this spacetime admits closed timelike curves localized in the interior of the blender at the ``beginning" of the universe. 
The nature of the blender automatically resolves the horizon problem, but Hawking's Theorem \ref{thm:hawking2} suggests that the metric cannot be modified to respect the strong energy condition while remaining geodesically complete. An interesting open question, Question \ref{quest:modify-blender}, is whether a modification could respect the more physical dominant energy condition while remaining complete.

In Section \ref{subsec:trumpet}, we discussed that one may avoid both the constraints of certain variants of Hawking's theorems and any causal pathologies by passing to the universal cover of the blender, the Eternal Trumpet spacetime. 
This spacetime is endowed with the same metric (\ref{eqn:blendermetric}), so it is still completely nonsingular and asymptotic to a radiation-dominated FLRW geometry, but it is also topologically trivial and globally hyperbolic, the strongest available notion of causally well-behaved. 
The effective cosmological slices in this model provide examples of general relativistic initial data sets which are arbitrarily small perturbations of FLRW initial data sets and which admit a past Cauchy development that is smoothly extendible to a complete and globally hyperbolic spacetime.
Known theorems provide no obstruction to there being a modification to the Eternal Trumpet's metric which respects the dominant energy condition while remaining timelike complete, suggesting Question \ref{quest:modify-trumpet}. If one seeks to satisfy the strong energy condition as well, known results force one to look between the blender and the trumpet, leading to Question \ref{quest:modify-between}.

As interesting as the present observations are, there remains a good deal of physics to be done to place any of these pictures of cosmology on comparable footing to (or interweave them with) the standard inflationary paradigm underlying $\Lambda$CDM, or to extract any unique observable signatures. Nonetheless, we hope to have demonstrated here that, singularity theorems notwithstanding, the immense breadth of general relativity affords a great variety of potential structures for the early universe that may be worthy of deeper consideration.

\section*{Acknowledgements}
The authors would like to thank David Garfinkle for providing helpful and illuminating commentary on the content of this manuscript, as well as the referees for making a number of suggestions improving its clarity and presentation.

\bibliographystyle{plain}
{\small \bibliography{references.bib}}

\end{document}